\documentclass[twocolumn,showpacs,prl,aps,superscriptaddress]{revtex4}
\usepackage{epsfig}

\usepackage{graphicx}
\usepackage{dcolumn}

\newcommand{\beq}{\begin{equation}}
\newcommand{\eeq}{\end{equation}}
\usepackage{graphicx}
\usepackage{amssymb}

\newcommand{\beqa}{\begin{eqnarray}}
\newcommand{\eeqa}{\end{eqnarray}}

 \newcommand{\la}{\langle}
\newcommand{\ra}{\rangle}
\newcommand{\var}{\varepsilon}

\def\arx#1{{ arXiv:} { #1}}

\def\oc#1{{ Opt.\ Commun.} {\bf#1}} \def\jmo#1{{ J.\ Mod.\ Opt.} {\bf#1}}
 
\def\pra#1{{ Phys.\ Rev. A\/} {\bf#1}} \def\prb#1{{ Phys.\ Rev. B\/} {\bf#1}}
 \def\prl#1{{ Phys.\ Rev.\
Lett.} {\bf#1}}
\def\sci#1{{ Science} {\bf#1}}

\begin{document}

\title{Effects of Quantum Error Correction on Entanglement Sudden Death}

\author{Muhammed  Y\"{o}na\c{c}}
\affiliation{Department of Electrical and Electronics Engineering, Zirve University, Gaziantep TURKEY 27260}
\author{J. H. Eberly}
\affiliation{Department of Physics and Astronomy, University of Rochester, Rochester NY 14627}

\date{\today}


\begin{abstract}

We investigate the effects of error correction on non-local quantum coherence as a function of time, extending the study by Sainz and Bj\"ork \cite{sainz-bjork}. We consider error correction of amplitude damping, pure phase damping and combinations of amplitude and phase damping as they affect both fidelity and quantum entanglement. Initial two-qubit entanglement is encoded in arbitrary real superpositions of both $\Phi$-type and $\Psi$-type Bell states. Our main focus is on the possibility of delay or prevention of ESD (early stage decoherence, or entanglement sudden death). We obtain the onset times for ESD as a function of the state-superposition mixing angle.  Error correction affects entanglement and fidelity differently, and we exhibit initial entangled states for which error correction increases fidelity but decreases entanglement, and vice versa.
\end{abstract}

\maketitle


\section{Introduction}


Entanglement is a condition shared by vectors in two or more separate vector spaces. As is well known quantum entanglement can provide an important resource for quantum computation and communication because it is a non-local effect, i.e., it can exist among multi-party systems completely disconnected from each other (see, e.g., Preskill \cite{preskill} and Nielsen and Chuang \cite{nielsen-chuang}).   The quantum entanglement in a multi-party system is a form of joint coherence that is subject to change whenever the system interacts with an external environment, typically becoming smaller.  Such coherence-diminishing interactions are almost always present, and since they can reduce or eliminate the usefulness of entanglement their effects are important to understand.

Environmental interactions can cause loss of coherence in single systems as well as joint systems, but entanglement is a more fragile coherence in some ways \cite{yu-eberly03}. A particular example is provided by what is labelled ESD (early stage decoherence or entanglement sudden death). This can take place in common physical contexts \cite{ESD} even under relatively benign dissipative interactions with the environment.

As the name implies, ESD is more destructive than single-system decoherence, and not in the sense that it is governed by a larger rate constant (which is usually not the case), but because ESD completely eliminates entanglement in a finite time (see manifestations in various model contexts: \cite{zyc, son-etal, daffer-etal, diosi, dodd-halliwell} and the review \cite{yu-eberly09}). This can happen while the local coherences of the same parties remain continuously non-zero under the same environmental interaction.  ESD is not well understood. Its onset is not predicted in any known way for a general entangled mixed state (however, see recent results \cite{qian-eberly12}). This is true even for the simplest case, qubit-qubit entanglement, which can be present in the smallest possible (4 $\times$ 4) joint density matrix.

Quantum error correction \cite{QEC} is a known weapon against single-system decoherence, but tests of its efficacy against multi-party decoherence and specifically against dissipation of quantum entanglement are rare \cite{sainz-bjork, wickert-etal}.  Here we report results of calculations that expand the domain for which results are available. We focus on the effects of quantum error correction on ESD under different noisy-environment interactions, using error correction algorithms that are convenient for this limited purpose. In practice a noisy channel is a dynamical process and its physical features are usually not fully understood or predictable, and so  are potentially a complication for error correction. Additionally, the multi-qubit logic states that are required to introduce needed redundancy multiply the overall resource requirements.

We particularly extend the quantum error correction study initiated by Sainz and Bj\"ork \cite{sainz-bjork} in which they showed that error correction can actually promote ESD in a bipartite system instead of preventing or delaying it. We present comparisons of time-dependent results for uncorrected and error-corrected  state evolution, focusing on two-qubit entanglement with concurrence as measure of entanglement. A comparison is made of concurrence with fidelity, also frequently used to describe multiparty coherence.  

The method that we use lets the error-channel in question affect our system for an amount of time before correction. That is, we assume that the qubit in question represents a stored resource between protocol-controlled gate operations and it is during this ``quiet" interval that most environment-induced errors occur. After this interval we isolate our system and subject it to error correction.  The later the isolation, the higher the chance of error occurrence.  The error correction methods we adopt encode the qubits that are to be protected into multi-bit logical codewords and they can perfectly correct the errors that effect only one of the bits in these codewords. The encoding operation is local.  Namely, given a pair of entangled and physically separated qubits, we apply encoding onto each of them at their own stations.  We will assume that this operation does not alter the entanglement between the qubits as we will not try to extract information on the states of the qubits during the encoding.  Finally, we will assume that the apparatus used in the error-correction is itself error-free.  

\begin{figure}[!h]
\includegraphics[width=8cm]{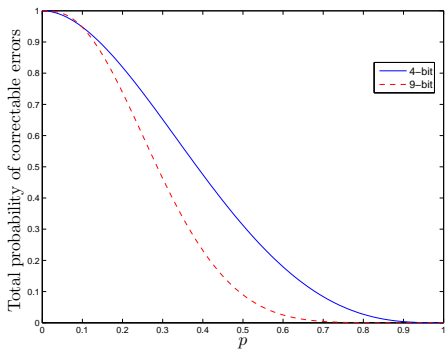}
\caption{The probability of occurrence of a correctable error vs. $p$, the probability of an error at a given bit. The effectiveness of the 9-bit code decreases substantially with respect to the 4-bit code for larger $p$.}   \label{4bit9bit}
\end{figure}

It is possible to use codes that correct multi-bit errors.  However, one has to increase the number of bits in the codewords in order to be able to correct higher order errors.  This in turn may increase the probability of uncorrectable errors depending on the probability of a single-bit error.  For instance, it has been shown that Shor's 9-bit code \cite{shor} can be used to correct amplitude damping errors that effect up to two bits in the codewords \cite{gottesman}.  Fig. \ref{4bit9bit} shows a comparison between this code and the $4$-bit code we use.  It is clear that unless the single-bit error probability, $p$, is smaller than about $0.1$, the success probability of the $4$-bit code is higher.  Since we will be dealing with the whole range of error probabilities, $0 \le p \le 1$, the single-bit error correction is more appropriate in our case.

\section{Amplitude Damping}


Amplitude damping here refers to the results generated by the interaction between a two-level quantum system (qubit) and a weakly coupled environment that is modeled as consisting of a continuum of harmonic oscillators. The boson modes of the electromagnetic vacuum are one example, but the model is practically general - any very weakly excited environmental interaction will behave as a harmonic oscillator. The damping effect is to drive the qubit to just one of its two possible states, thus removing any possibility of coherent superposition of the states. The Hamiltonian representing the interaction can be written,

\beq
H^{AD}_I=\sum_k (g_k\sigma_+ a_k + g_k^*\sigma_- a_k^{\dagger}),
\eeq
where $\sigma_\pm$ are the raising and lowering Pauli operators for the atom and $a_k$, $a_k^{\dagger}$ are the annihilation and creation operators for the $k^{th}$ mode of the environment. The strength of interaction between the atom and each mode is determined by the parameter $g_k$.  Under the Born-Markov approximation \cite{carmichael}, the master equation for the time evolution of the single-qubit density matrix is given in the interaction picture by,
\beq
\dot{\rho}=\Gamma_{AD}(2\sigma_-\rho \sigma_+ - \sigma_+\sigma_-\rho - \rho\sigma_+\sigma_-),
\eeq
where $\Gamma_{AD}$ is the rate of damping which depends on the spectral properties of the environment and the interaction parameter.  The Kraus operators representing this Markovian evolution are,
\beqa
E_0 = \left(
  \begin{array}{cc}
    1 & 0 \\
   0 & \sqrt{1-p_{ad}} \\
  \end{array}
\right),\ \
E_1 = \left(
  \begin{array}{cc}
    0 & \sqrt{p_{ad}} \\
   0 & 0 \\
  \end{array}
\right),
\eeqa\label{ADKraus}
where $p_{ad} = 1-e^{-2\Gamma_{AD} t}$.  The environment-induced evolution implied by Eq. 2 can then be obtained using the Kraus operators:
\beq
\rho_\var(t)=\sum_k E_k \rho(0) E_k^{\dagger},
\eeq
where the subscript $\var$ indicates error-channel evolution. This Kraus representation ensures that the $AD$ channel has no effect on the $|0\rangle$ state while it changes $|1\rangle$ into $|0\rangle$ with probability $p_{ad}$ and leaves it unchanged with probability $1 - p_{ad}$. This can be translated into a state amplitude-changing map for the joint atom-boson mode state:
\beqa \label{ADmap}
|0\ra|vac\ra & \to & |0\ra|vac\ra \nonumber \\
|1\ra|vac\ra & \to & \sqrt{1-p_{ad}}\ |1\ra|vac\ra + \sqrt{p_{ad}}\ |0\ra|\phi\ra,
\eeqa
where $|\phi\ra$ is the one-boson state created in the process.

The error correction process works in four steps: First, the state is encoded using codewords that consists of multiple bits. Second, the encoded state is exposed to the error channel.  Third, the system is isolated and a syndrome measurement is performed in order to find out the type of error that occurred.  Finally, conditional on the result of the syndrome measurement, a recovery operation is performed on the system in order to (partially) restore the original state.  Thus, given the initial encoded state $\rho_0$, and its error-modified form $\rho_\var$, the recovered state, $\rho_R$ is given by,
\beq
\rho_R=\sum_k R_k M_k \rho_\var M_k^\dagger R_k^\dagger,
\eeq
where $M_k$ is the syndrome measurement operator that detects the $k^{th}$ type error and $R_k$ is the corresponding recovery operator.  We describe the method more fully in Sec. \ref{QEC method} at the end.

Now we address non-local coherence under error-channel evolution and correction. Following \cite{sainz-bjork}, we apply this procedure to a pair of entangled (but possibly remotely located) qubits, both interacting only with their local environments, taken to be identical and exposing them to the same type of amplitude-damping error.  First we test the entanglement of a generalized Bell-type state $|\Phi\ra$, which is a superposition of the usual $|\Phi^{\pm}\ra$ states, known to be ESD-susceptible, to represent the initial atomic state:

\beqa
|\Phi\rangle &=& \cos\delta|\Phi^+\ra + \sin\delta|\Phi^-\ra \nonumber \\
&=& \cos\alpha|11\rangle + \sin\alpha|00\rangle,
\eeqa
where $\alpha + \delta = \pi/4$. Since the $|11\ra$ and $|00\ra$ states fit the error correction steps better than the $|\Phi^{\pm}\ra$ states, we will call $\alpha$ the mixing angle for the superposition.

We then calculate the fidelity and degree of entanglement of the corrected state and compare each with the values for the uncorrected one.  We measure entanglement with concurrence \cite{wootters}, and under the environment's amplitude-damping influence we find for concurrence and fidelity:
\beqa\label{CPhiAD}
C_\Phi(\rho_\var) & = & \hbox{max}\Big[0,2(1-p_{ad})|\cos\alpha|\nonumber\\
& \times & \Big(|\sin\alpha|-|\cos\alpha|p_{ad}\Big)\Big], \\
F_\Phi(\rho_\var) & = & \la\Phi(0)|\rho(t)|\Phi(0)\ra\nonumber\\
 &=& 1-2p_{ad}\cos^2\alpha+p_{ad}^2\cos^2\alpha.
\eeqa
There is no general analytical formula for $C_\Phi(\rho_R)$ and $F_\Phi(\rho_R)$, the concurrence and fidelity of the state after the error correction.  However, it is possible to find such formulas with a small $p_{ad}$ approximation \cite{sainz-bjork}.

\begin{figure}[!t]
\includegraphics[width=9cm]{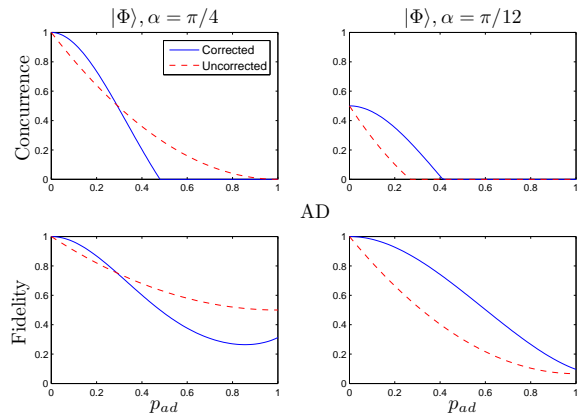}
\caption{Concurrence (top) and fidelity (bottom) vs. $p$ for $|\Phi\rangle$ under amplitude damping, showing the different effects of error correction depending on $\alpha = \pi/4$ (left) and $\alpha = \pi/12$ (right).  The dashed line represents $\rho_\var$, the error-affected but ``uncorrected'' state, and the solid line represents $\rho_R$ which is the mapping of $\rho_\var$ under the error correction operation, thus referred to as the ``corrected'' state. }\label{CF-PhiAD}
\end{figure}

Fig. \ref{CF-PhiAD} shows the results for maintenance of concurrence and fidelity for two values of the initial superposition mixing angle $\alpha$. Time is represented parametrically by $p_{ad} = 1 - e^{-\Gamma_{AD}t}$, ranging from 0 to 1 (t=0 to t=$\infty$).  The figures should be interpreted as follows: Our qubits start from a pure state and evolve under the error channel.  The dashed lines represent this evolution (we refer to them as ``uncorrected").  For each time point, the corresponding point on a solid line represents the state that is reached from the dashed line after error correction.

Several features are worth note in Fig. \ref{CF-PhiAD}. For example, fidelity is confirmed as a bad substitute for concurrence, i.e., it is not a reliable measure of entanglement. Fidelity is finite for all time, whereas entanglement vanishes at a finite time in all cases except the uncorrected pure Bell state ($\alpha = \pi/4$). The remarkable feature first noted by Sainz and Bj\"ork \cite{sainz-bjork} is shown in the top left panel, where correction of the amplitude damping effect actually initiates ESD rather than preventing it. However, one can also see that a departure from pure Bell status ($\alpha = \pi/12$) leads to more rapid ESD, but the error correction in that case is somewhat effective, not preventing ESD but at least delaying it.  We can also see that the time of error correction must be chosen carefully. On the top left panel, it can be seen that around $\gamma_{AD}=0.2$ the error correction maps the error-affected state into one with a higher concurrence.  However, if the error correction is instead applied at $\gamma_{AD}=0.4$ the state is mapped into one with a lower concurrence.  In fact, it is evident from the same figure that there is no use in applying error correction once time reaches around $\gamma_{AD}=0.3$.  In order to increase the entanglement, correction must be applied before this point.

\begin{figure}[!t]
\includegraphics[width=8cm]{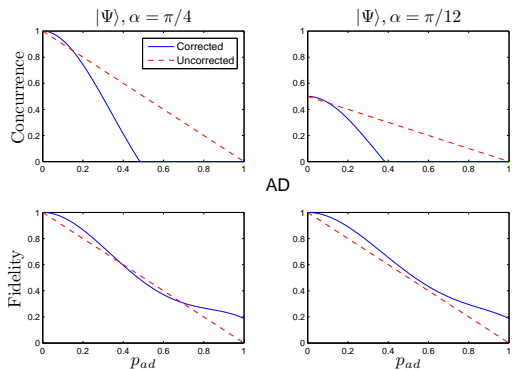}
\caption{Concurrence (top) and fidelity (bottom) vs. $p$ for the $|\Psi\rangle$ superposition under amplitude damping, showing the effects of error correction in its dependence on two superposition angles, $\alpha = \pi/4$ (left) and $\alpha = \pi/12$ (right).  Error correction is of little consequence for fidelity. It provides positive assistance to entanglement, but only for a small fraction of the amplitude damping halflife, after which it quickly induces ESD.}\label{CF-PsiAD}
\end{figure}

For the other Bell state we write its generalized superposition as
\beqa  \label{CPsiAD}
|\Psi\ra & = & \cos\delta|\Psi^+\ra + \sin\delta|\Psi^-\ra \nonumber \\
& = & \cos\alpha|10\rangle + \sin\alpha|01\rangle,
\eeqa
where again $\alpha + \delta = \pi/4$. We also calculate the fidelity and degree of entanglement of the corrected states and compare them in Fig. \ref{CF-PsiAD} with those of the uncorrected ones.  Under the environment's influence we find:
\beqa
C_\Psi(AD)  &=&  2|\sin\alpha\cos\alpha|(1-p_{ad}),\ {\rm and} \\
F_\Psi(AD) &=& 1-p_{ad},
\eeqa
so that fidelity is independent of the mixing angle, and entanglement of the state remains finite for all time ($p_{ad} < 1$) in the absence of testing for error.

However, Fig. \ref{CF-PsiAD} shows that when error correction is instituted (and amplitude errors are surely injected by the damping), one consequence is improvement in concurrence (top panels) for a very short period, but quickly such improvement disappears and a strong ESD event is induced by the error correction. As in the $\Phi$ state case, the response of fidelity to error correction again has little relation to the response of entanglement.


\section{Phase Damping (PD)}


As a second error channel, we examine pure phase damping (PD), which in some situations can be more rapid than amplitude damping. It can be interpreted as arising from rapid stochastic shifts of the qubit transition frequency, which has a clear classical analog (see e.g., \cite{yu-eberly03, yu-eberly06}), and can be easily extended beyond the Markov limit \cite{yu-eberly10}. Using bosonic reservoir language, in pure phase damping a boson is created or annihilated in the environment while the energy of the atom is unchanged; thus it can also be interpreted as arising from weak scattering processes. The interaction part of the Hamiltonian engages the qubit's inversion, and is given by,
\beq
H_I^{PD}=\sigma_z\sum_k(h_k a_k + h_k^* a_k^{\dagger}).
\eeq

The master equation for phase damping in the interaction picture can be written as,
\beq
\dot{\rho}=\Gamma_{PD}(\sigma_z \rho \sigma_z - \rho),
\eeq
where $\Gamma_{PD}$ is the phase-damping parameter.  Only the off-diagonal elements are affected:
\beqa
\rho_{11}&=&\rho_{11}(0)\ {\rm and}\ \rho_{22} = \rho_{22}(0)\nonumber\\
\rho_{12}&=&\rho_{12}(0)e^{-2\Gamma_{PD} t}\ {\rm and}\ \rho_{21} = \rho_{12}^*,
\eeqa
and the Kraus operators for this evolution are,
\beqa
E_0 = \left(
  \begin{array}{cc}
    1 & 0 \\
   0 & \sqrt{1 - p_{pd}} \\
  \end{array}
\right),\ \
E_1 = \left(
  \begin{array}{cc}
    0 & 0 \\
   0 & \sqrt{p_{pd}} \\
  \end{array}
\right),
\label{PD-kraus}
\eeqa
where $p_{pd} = 1 - e^{-2\Gamma_{PD} t}$.

Once again, this code can fully correct the first order errors, where one bit is flipped. In this case, the concurrences and fidelities for the corrected and uncorrected states are equal to each other:
\beqa
C_{\Phi}(\rho_\var) &=& C_{\Psi}(PD) \nonumber \\
&=& 2|\sin\alpha\cos\alpha|(1-\gamma_{PD}) \\
F_{\Phi}(\rho_\var) &=& F_{\Psi}(PD) \nonumber \\
&=& 1- 2\gamma_{PD}\sin^2\alpha\cos^2\alpha.
\eeqa

\begin{figure}[!h]
\includegraphics[width=8cm]{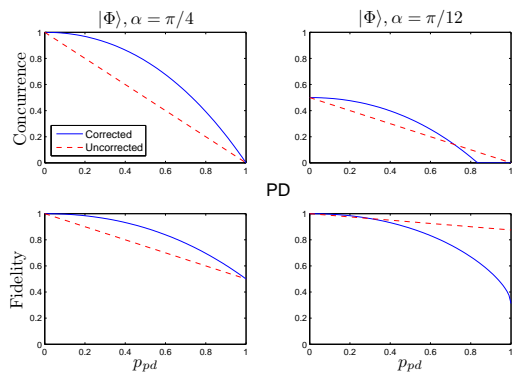}
\caption{Concurrence (top) and fidelity (bottom) are plotted vs. $p_{pd}$ under phase damping for $\alpha = \pi/4$ (left) and $\alpha = \pi/12$ (right). As before, the effect of elementary error correction is shown in blue. The same figures apply to both $|\Phi\ra$ and $|\Psi\ra$.}\label{CF-PhiPD}
\end{figure}

Note that uncorrected concurrence is above zero for all times ($p_{pd} < 1$), only vanishing at $p_{pd} = 1 (t = \infty$). However, the correction algorithm makes a difference.  Fig. \ref{CF-PhiPD}, which serves for both $|\Phi\ra$ and $|\Psi\ra$, shows that error correction (solid line) counteracts phase damping and maintains the entanglement of a single Bell state ($\alpha = \pi/4$ for either $\Phi$ or $\Psi$) above the uncorrected value for all time, but when applied to Bell superposition states ($\alpha \neq \pi/4$), can give rise to ESD, as in the upper right panel. Similarly, fidelity of a single Bell state is well protected for arbitrarily long times, but is very poorly protected for a Bell superposition even for short times, as seen in the bottom right panel.

\section{Relative Effect (AD and PD)}

\begin{figure}[!b]
\includegraphics[width=9cm]{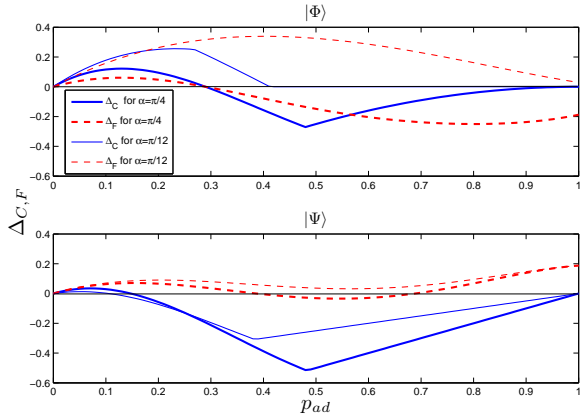}
\caption{The differences between the concurrences and fidelities with or without correction for $|\Phi\ra$ and $|\Psi\ra$ under AD.  The results are plotted for two different values of $\alpha=\pi/4$ and $\alpha=\pi/12$.}\label{DeltaCF-PsiAD}
\end{figure}

Fig. \ref{DeltaCF-PsiAD} shows the effect of the error correction on the concurrence and fidelity more clearly.  Here we introduce the differences between corrected and uncorrected quantities for a direct indication of correction effectiveness.
\beqa
\Delta_C &=& C_{\Psi}(\rho_R)-C_{\Psi}(\rho_\var) \nonumber \\
\Delta_F &=& F_{\Psi}(\rho_R)-F_{\Psi}(\rho_\var).
\eeqa
The interesting point is that there are regions for both values of superposition angle $\alpha$ where $\Delta_F$ is positive while $\Delta_C$ is negative.  In other words it is possible that the error correction increases the fidelity while it decreases the entanglement.  We will see in the next section that the opposite can also be true.

\begin{figure}[!t]
\includegraphics[width=9cm]{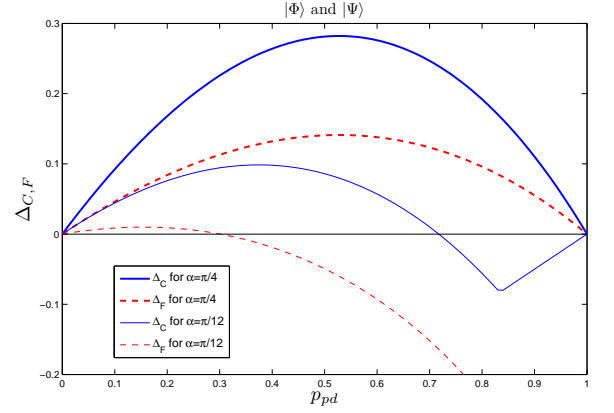}
\caption{The differences between the concurrences and fidelities with or without correction under the PD.  Both $|\Phi\ra$ and $|\Psi\ra$ have the same dynamics under PD. }\label{DeltaCF-PhiPD}
\end{figure}

From the data displayed above we can easily extract the absolute differences that are found between corrected and uncorrected records. These differences are shown in Figs. \ref{DeltaCF-PhiPD} and \ref{DeltaCF-PsiAD} for amplitude damping and phase damping respectively.  Here we are defining differences as
\beqa
\Delta_C&=&C_{\Psi}(\rho_R)-C_{\Psi}(\rho_\var) \nonumber \\
\Delta_F&=&F_{\Psi}(\rho_R)-F_{\Psi}(\rho_\var).
\eeqa

The interesting point is that there are regions for both values of $\alpha$ where $\Delta_F$ is positive while $\Delta_C$ is negative.  In other words it is possible that the error correction increases the fidelity while it decreases the entanglement.

Fig. \ref{DeltaCF-PhiPD} shows the effect of error correction on the concurrence and fidelity of the $|\Phi\ra$ state for two different $\alpha$ values.  The right panel has regions where $\Delta_C$ is positive while $\Delta_F$ is negative.  In other words, error correction may lead to an increase in entanglement and a decrease in fidelity.

\section{Combined Noise (AD and PD)}


A remaining task is to examine combined AD and PD errors, since in general we would expect to have both of these error mechanisms present together to some extent in a physical system.  The interaction part of the Hamiltonian in the combined noise case is

\beqa
H_I^{AD+PD} &=& \Gamma_{AD}(2\sigma_-\rho \sigma_+ - \sigma_+\sigma_-\rho - \rho\sigma_+\sigma_-)\nonumber \\
&+& \Gamma_{PD}(\sigma_z \rho \sigma_z - \rho),
\eeqa
where $\Gamma_{AD}$ and $\Gamma_{PD}$ are the damping parameters for the AD and PD respectively when they are applied separately.  Clearly, the relation of the two damping rates may be important, and we will designate it by the ratio:
\beq \label{kappadef}
\kappa = \Gamma_{PD}/\Gamma_{AD}.
\eeq
The solution for qubit evolution is given by,

\beqa
\rho_{11}&=&1-\rho_{22}(0)e^{-2\Gamma_{AD} t}\nonumber\\
\rho_{12}&=&\rho_{12}(0)e^{-\Gamma_{AD}-2\Gamma_{PD} t}\nonumber\\
\rho_{21}&=&\rho_{12}^*\nonumber\\
\rho_{22}&=&\rho_{22}(0)e^{-2\Gamma_{AD} t}.
\eeqa
A two-operator Kraus sum for this type of combined error does not exist.  Instead we can find a three-operator sum with the elements,

\beqa
E_0&=&\left(
  \begin{array}{cc}
    1 & 0 \\
   0 & \sqrt{(1-p_{ad})(1-p_{pd})} \\
  \end{array}
\right),\nonumber\\
E_1&=&\left(
  \begin{array}{cc}
    1 & 0 \\
   0 & \sqrt{(1-p_{ad})p_{pd}} \\
  \end{array}
\right),\nonumber\\
E_2&=&\left(
  \begin{array}{cc}
    0 & \sqrt{p_{ad}} \\
   0 & 0 \\
  \end{array}
\right).
\eeqa
where $p_{ad}$ and $p_{pd}$ have already been defined.  In order to make the AD and PD parts of the sum more transparent, we apply a unitary transformation to these elements to turn them into,

\beqa
E_0^{'}&=&\frac{x}{\sqrt{x^2+y^2}}\left(
  \begin{array}{cc}
    1 & 0 \\
   0 & x+\frac{y^2}{x}-1 \\
  \end{array}
\right),\nonumber\\
E_1^{'}&=&\frac{y}{\sqrt{x^2+y^2}}\left(
  \begin{array}{cc}
    1 & 0 \\
   0 & -1 \\
  \end{array}
\right),\nonumber\\
E_2^{'}&=&\left(
  \begin{array}{cc}
    0 & \sqrt{p_{ad}} \\
   0 & 0 \\
  \end{array}
\right),
\eeqa
where,
\beqa
x&=&1+\sqrt{(1-p_{ad})(1-p_{pd})},\nonumber\\
y&=&\sqrt{(1-p_{ad})p_{pd}}.
\eeqa
$E_1^{'}$ and $E_2^{'}$ are the error operators for the PD and AD respectively. $E_0^{'}$ represents the no-error case for small $p_{ad}$ and $p_{pd}$.

\begin{figure}[!t]
\begin{center}
\epsfig{file=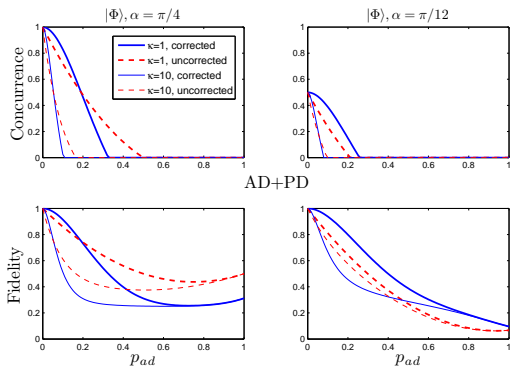, width=8cm}
\end{center}
\caption{Concurrence and fidelity  are plotted vs. $p_{ad}$ for $|\Phi\rangle$ under the combined noise for $\alpha = \pi/4$ (left) and $\alpha = \pi/12$ (right) using $\kappa=1$ and $\kappa=10$ ($\kappa = \Gamma_{PD}/\Gamma_{AD}$).  For $\kappa=10$ entanglement of both corrected and uncorrected states vanish before $p_{ad} = 0.4$ whereas fidelity always stays above zero.} \label{ad+pd-phi}
\end{figure}

\begin{figure}[!b]
\begin{center}
\epsfig{file=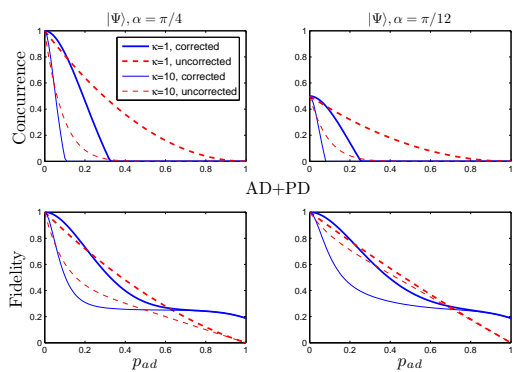, width=8cm}
\end{center}
\caption{Concurrence (top) and fidelity (bottom) are plotted vs. $p_{ad}$ for $|\Psi\rangle$ under the combined noise for $\alpha = \pi/4$ (left) and $\alpha = \pi/12$ (right) using two different $\kappa$ values, $\kappa=1$ and $\kappa=10$.  Increasing $\kappa$ decreases the ESD onset time. Also, as for the AD and PD cases, error-correction leads to smaller ESD onset times.   The features seen in Fig. 5 are observed here as well. } \label{ad+pd-psi}
\end{figure}

We will use the [5,1] code proposed by Laflamme \emph{et al.} \cite{laflamme} for the error correction (see Sec. VII).  As in \cite{sainz-bjork}, we will ignore the distortion caused by the second diagonal element of $E_0^{'}$ and assume that the no-error subspace is spanned by $|0_L\ra$ and $|1_L\ra$.  The one-bit PD (AD) error subspace is spanned by the vectors each of which results from the application of $\var_{1}$($\var_{2}$) on the relevant qubit of $|0_L\ra$ and $|1_L\ra$  and identity operators on the rest, where,
\beqa
\var_1&=&\left(
  \begin{array}{cc}
    1 & 0 \\
   0 & -1 \\
  \end{array}
\right),\nonumber\\
\var_2&=&\left(
  \begin{array}{cc}
    0 & 1 \\
   0 & 0 \\
  \end{array}
\right).
\eeqa
In the end we get $22$ orthogonal vectors, each corresponding to a different type of one-bit error (or a no-error) originated in $|0_L\ra$ or $|1_L\ra$.  The remaining $10$ vectors that are necessary to span the $32$-dimensional codeword space correspond to the errors that cannot be corrected and if the syndrome measurement points to one of them we project our original qubit onto $(|0\ra\la0|+|1\ra\la1|)/2$.  The results are given in Fig. \ref{ad+pd-phi} and Fig. \ref{ad+pd-psi}.

The concurrences and fidelities for the uncorrected states are given by,

\beqa
C_\Phi(\rho_\var) &=& \hbox{max}\Big[0, 2(1 - p_{ad})|\cos\alpha|\nonumber \\
&&\times\Big(|\sin\alpha|(1 - p_{pd}) - |\cos\alpha|p_{ad}\Big)\Big],\\
F_\Phi(\rho_\var) & = & 1 - 2p_{ad}\cos^2\alpha + p_{ad}^2\cos^2\alpha \nonumber\\
&& -2p_{pd}(1-p_{ad})\sin^2\alpha\cos^2\alpha \\
C_\Psi(\rho_\var) & = & 2|\sin\alpha\cos\alpha|(1 - p_{ad})(1- p_{pd}) \\
F_\Psi(\rho_\var)& = & 1 - p_{ad} - 2p_{pd}(1 - p_{ad})\sin^2\alpha\cos^2\alpha.
\eeqa


\section{ESD Onset Times}


The factors that govern or control ESD are still generally unknown, particularly in the case of initially mixed state evolution. One feature that would be valuable to have under control is the onset time for ESD. To our knowledge, no attempt to examine the onset time systematically has been made. A recent exception is found in \cite{qian-eberly12}.

We now let $T$ be the time when $C$ reaches zero, i.e., $T$ is the ESD onset time. Then from (\ref{CPhiAD}) we obtain the ESD onset value of $p_{ad}$ for $|\Phi\rangle$ as
\beq
p_{ad}(T) = |\tan\alpha|.
\eeq

\begin{figure}[!h]
\begin{center}
\epsfig{file=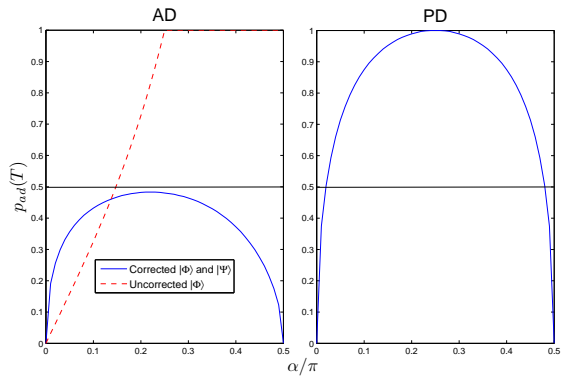, width=8cm}
\end{center}
\caption{ESD onset time, measured by the relevant $p$, is plotted vertically vs. the Bell superposition parameter $\alpha$ for AD (left) and PD (right). Uncorrected $|\Psi\ra$ does not show ESD under either AD or PD, and uncorrected $|\Phi\ra$ does not show ESD under PD.  As a result, these cases are not plotted.} \label{ESDtime}
\end{figure}

However, some results of this kind can be obtained from an extension of our examinations of concurrence for both uncorrected and corrected evolution. We have computed the evolution of concurrence, under the correction algorithms used above, for a wide range of $\alpha$ values. For essentially every $\alpha$ there is found a finite onset time for ESD. In  Fig. \ref{ESDtime} we show the way the ESD onset time, measured by the appropriate $p$, depends on $\alpha$.

The dashed line in the left panel of Fig. \ref{ESDtime} shows that uncorrected concurrence is preserved for long times if the initial state was nearly a single Bell state ($\alpha \approx \pi/4$). The solid line shows that application of the error correction algorithm rapidly induces ESD for all values of $\alpha$.  In the right panel the solid line shows that phase damping is much more effectively forestalled by the correction algorithm used. Only values of $\alpha$ substantially different from $\pi/4$ lead to a short ESD time under phase damping.

It is evident that $|\Psi\rangle$ combination does not show ESD under combined AD and PD effect.  For $|\Phi\rangle$, the ESD onset time is determined by,

\beq
\frac{p_{ad}(T)}{1 - p_{pd}(T)}=|\tan\alpha|.
\eeq
This should be compared with the ESD onset for AD which is given by,

\beq
p_{ad}(T)=|\tan\alpha|.
\eeq
Since for any given time,

\beq
\frac{p_{ad}}{1-p_{pd}} \ge p_{ad},
\eeq
Eq. 38 is satisfied at an earlier time than Eq. 39 (except for $\alpha=0$).  Thus, $T$ decreases with the introduction of the PD to an AD-only system.  The uncorrected states do not give ESD in a PD-only system.  The introduction of even a slight amount of AD, however, leads to a finite $p_{ad}/(1 - p_{pd})$ which means ESD will set in after a finite amount of time when AD is introduced into a PD-only system.

\begin{figure}[!t]
\begin{center}
\epsfig{file=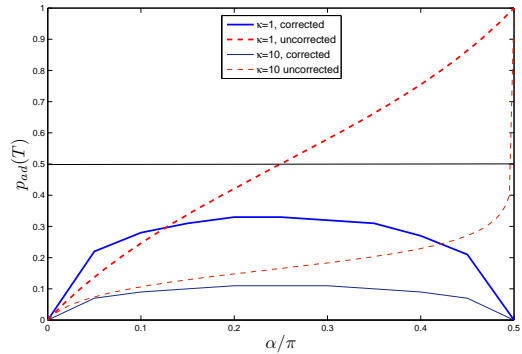, width=8cm}
\end{center}
\caption{ESD onset time for $|\Phi\ra$, measured by $p_{ad}$ is plotted vertically vs. the Bell superposition parameter $\alpha$ for the combined noise using two different $\kappa$ values, $\kappa =1$ and $\kappa =10$.  ESD onset time decreases with increasing $\kappa$.  For $\kappa=10$, uncorrected state has a larger ESD time than the corrected state for the whole range of $\alpha$. Corrected $|\Psi\ra$ gives the same results.  Uncorrected $|\Psi\ra$ does not show ESD.} \label{ad+pd-ESDtime}
\end{figure}


\section{Sketch of Method } \label{QEC method}


The main challenge in the error correction process is finding a suitable code.  The $[4,1]$ code given by Leung \emph{et al.} \cite{leung} is an example of such a code, and we have used it for amplitude damping.  It encodes the two qubit states into two logical states that each live in a $4$-dimensional Hilbert space.  Each qubit is encoded using four bits:
\beqa
|0\rangle_L&=&(|0000\rangle + |1111\rangle)/\sqrt 2,\nonumber\\
|1\rangle_L&=&(|0011\rangle + |1100\rangle)/\sqrt 2,
\eeqa
and the scheme can fully correct only the errors that come from a single AD error which works like a one-way bit-flip, annihilating $|0\ra$ while changing $|1\ra$ into $|0\ra$.  The AD channel is then applied separately onto the four code bits which in itself gives an error operation consisting of $16$ operators ($2$ operators for each qubit).

For example, suppose that we have a single-qubit state given by $a|0\ra + b|1\ra$ which is encoded as $a|0\ra_L + b|1\ra_L$ and exposed to the AD error.  Assume that, during the error process, an error occurs only at the first qubit in the code.  Following the notation used by Sainz and Bj\"ork \cite{sainz-bjork}, the states that are reached from the codewords under this error are:
\beqa
|r_{10}\ra&=&|0111\ra \nonumber\\
|r_{11}\ra&=&|0100\ra,
\eeqa
where the first index 1 shows the number of the qubit that is affected by the error and the second index 0 or 1 shows which codeword includes the affected qubit.  Note that the two states are orthonormal.  This is an essential property which is necessary to have, to be able to distinguish the source of the error.  It is also essential that the states for different types of errors are orthonormal as well.  This ensures that one can distinguish between different error types upon a syndrome measurement and thus can recover the initial state correctly.

Returning to our example, the syndrome measurement operator for this error type is given by,
\beq
M_1=|0111\ra\la0111| +|0100\ra\la0100|,
\eeq
and the corresponding recovery operator is given by,
\beq
R_1=|0\ra_L\la0111| + |1\ra_L\la0100|,
\eeq
so it can be shown that,
\beq
R_1 M_1 (a|0\ra_L + b|1\ra_L)=a|0\ra_L + b|1\ra_L.
\eeq
Thus the original state is retrieved.  The other one-bit errors and the no-error case are recovered similarly.  It is shown in \cite{sainz-bjork} that the vectors that are reached from the codewords under these errors are all orthonormal, so these errors can be distinguished from each other and corrected.

The main problem with this procedure is that the correctable errors constitute only a fraction of all the errors.  For instance, in this case, there are $5$ types of correctable errors but in total we have $16$ error types.  Some of these errors are not detected by this procedure at all and the ones that are detected get mixed up with one of the five errors that can be corrected.  For instance the error type where the first, third and fourth qubits are damped lead to the vector $|0100\ra$ when applied to $|0\ra_L$.  However, as we have seen above, the same vector is also reached from $|1\ra_L$ upon application of the error type where only the first qubit is damped.  Thus, our procedure treats this error as if it stemmed from $|1\ra_L$ and recovers it as such while it should have actually given $|0\ra_L$.  The probability of having these high-order errors is small if time (or equivalently the damping parameter $p_{ad}$) is small.  Thus, the correction mechanism gives good results for small times.  However, this weakness becomes more visible as time increases and the correction eventually breaks down, meaning that leaving the state uncorrected gives better longer-term entanglement and fidelity values.

As for the PD case, a unitary recombination of $E_0$ and $E_1$ in Eq. \ref{PD-kraus} gives a new and equivalent set of operation elements for phase damping [see Theorem 8.2 of \cite{nielsen-chuang}],

\beqa
E_0^{'} = \sqrt{\beta}\left(
  \begin{array}{cc}
    1 & 0 \\
   0 & 1 \\
  \end{array}
\right),\ \
E_1^{'} = \sqrt{1-\beta}\left(
  \begin{array}{cc}
    1 & 0 \\
   0 & -1 \\
  \end{array}
\right),
\eeqa
where
\beq
\beta = \frac{1+\sqrt{1 - p_{pd}}}{2}.
\eeq
One can see that the effect of the error operator $E_1^{'}$ above is to change the state $|+\rangle=(|0\rangle + |1\rangle)/\sqrt{2}$ into the state $|-\rangle=(|0\rangle - |1\rangle)/\sqrt{2}$ and vice versa.  There is a three-bit error correcting code for the type of errors that flip the two states into each other \cite{nielsen-chuang}, in which our qubit states are encoded as
\beqa
|0\rangle_L=|---\rangle,\nonumber\\
|1\rangle_L=|+++\rangle.
\eeqa
The states reached by a single-bit error from these codewords are orthonormal to each other and to the codewords.  Thus, the code can accurately detect and recover from single-bit errors.  However, it is not reliable for correcting higher order errors.  For instance, an error that affects the first two qubits of $|0\ra_L$ is detected to be an error that affects the third qubit of $|1\ra_L$ and thus the state is recovered as $|1\ra$ instead of $|0\ra$.

In the case of combined AD-PD damping we have used the [5,1] code proposed by Laflamme \emph{et al.} \cite{laflamme} for the error correction.  The codes that we have used for AD and PD were specifically designed to cope with one-bit errors of their respective error type.  This code, on the other hand, can correct general one-bit errors.  In our case, it can be used to correct the errors where only one of the qubits in the codewords is effected by either AD or PD.  Errors that effect more than one qubit cannot be corrected with this code, as it was the case for the codes given in previous sections. The logical bits of the code are given by,

\beqa
|0\rangle_L&=&\Big(|00000\ra + |11100\ra - |10011\ra - |01111\ra\nonumber\\
 &&+ |11010\ra + |00110\ra + |01001\ra + |10101\ra\Big)/2\sqrt{2},\nonumber\\
|1\rangle_L&=&\Big(-|00011\ra + |11111\ra - |10000\ra + |01100\ra\nonumber\\
 &&+ |11001\ra - |00101\ra - |01010\ra + |10110\ra\Big)/2\sqrt{2}.\nonumber\\
\eeqa

There are $10$ types of error that this code can correct: Half of them coming from the application of a single AD error and the other half from the application of a single PD error on each qubit in the logical bits.  One can show that the states reached from the codewords through these errors are orthonormal to each other and to the codewords.  As a result, given that one of these errors occurred, a syndrome measurement can determine exactly which has occurred and the initial state can be restored through the application of the proper recovery operation.  However, as in the [4,1] code, higher order errors get mixed up with the correctable errors and this leads to a decrease in the efficiency of the code as the error probability increases.


\section{Summary and Conclusions}


We have examined the dynamics of two-qubit systems under amplitude and phase damping, where the noises are both applied separately and as a combination, with and without the assistance of error-correction procedures. The examples are not complicated, and the error correction algorithms are generically specific but not sophisticated. However, we believe they are general enough that conclusions can be drawn about preservation of prepared quantum coherence in arbitrary superpositions of generic $|\Phi\ra$ and $|\Psi\ra$ Bell states. Coherence can be measured in several ways, and we employ both fidelity and concurrence.

It is well-known that concurrence \cite{wootters} is a completely reliable measure (both necessary and sufficient) to evaluate the degree of entanglement that may be present in a two-qubit state. Fidelity is examined because it is sometime used as a substitute in place of concurrence, and implications about entanglement are sometimes drawn from it. However, it is not a measure of entanglement, and not a reliable substitute. As we have seen, it suffers its own difficulties under error correction, not correlated in an obvious way with those for entanglement.

The main conclusions are straightforward: As Sainz and Bj\"ork \cite{sainz-bjork} first discovered for amplitude damping, as we have confirmed in Fig. \ref{CF-PsiAD}, an error-correction process can itself give rise to ESD, i.e., kill entanglement completely. This clearly establishes that error correction does not always preserve entanglement. Additionally, as in the combined AD and PD case, it can decrease the ESD onset time significantly, depending on the strengths of the AD and PD damping parameters.

We do see that error correction can act positively with good success in some cases. Unfortunately, a systematic discrimination of cases is still lacking. This is illustrated for pure phase damping in Fig. \ref{CF-PhiPD}, where concurrence is strongly protected for all $p_{ad}$ values in the case $\alpha = \pi/4$, while it is not for $\alpha = \pi/12$, for which it goes into ESD after 2-3 half-lives of the dissipation.  Second, a finite fidelity does not necessarily indicate a finite entanglement, which is shown by amplitude damping for both $\alpha$ values tested.  Overall, it seems safe only to say that error correction may affect entanglement and fidelity in opposite ways, increasing one while decreasing the other.

One should note that quantum error correction is not the only method that can be used for protection against interactions with the environment, especially for multi-partite systems.  For instance, in a recent study, Chaves \emph{et al.} \cite{chaves} have shown that for a system with multiple parts, a local transformation in the initial state might be sufficient to preserve non-local effects, such as the entanglement.   It can be shown that for a non-maximally entangled two-qubit initial state under the AD channel, a local unitary transformation may increase the ESD onset times and even prevent it altogether (the method does not result in an improvement for a Bell state as the state remains unaltered under local unitary transformations).  It would be very interesting to examine the combined effect of such a transformation and the quantum error correction in a future work.  

Acknowledgements: We have benefitted from converations and discussions with S. Agarwal, G. Bj\"ork, C. Broadbent, J.G. Dressel, A.N. Jordan, and X.F. Qian; and financial support is acknowledged from the DARPA QuEST program and AFOSR award FA9550-12-1-0001.



\begin{thebibliography}{99}

\bibitem{sainz-bjork} I. Sainz and G. Bj\"{o}rk, \pra {77}, 052307 (2008), and Int. J. Quantum Inf. {\bf 7}, S245-255 (2008).

\bibitem{preskill} J. Preskill, {\it Quantum Information and Computation}, see Caltech Lecture Notes, http://theory.caltech.edu/~preskill/ph229/.

\bibitem{nielsen-chuang} M. Nielsen and I. Chuang, {\em Quantum Computation and Quantum Information} (Cambridge University Press, 2000).

\bibitem{yu-eberly03}  T.Yu and J.H. Eberly, \prb {68}, 165322 (2003).

\bibitem{ESD} A theoretical treatment of ESD in spontaneous emission is given in: T. Yu and J. H. Eberly, \prl {93}, 140404 (2004), and an experimental observation in a photonic context is reported in  M.P. Almeida et al., \sci {316}, 579 (2007). See J.H. Eberly and T. Yu, \sci{316}, 555 (2007) for comments.

\bibitem{zyc} K. Zyczkowski, P. Horodecki, M. Horodecki, and R. Horodecki, \pra {65}, 012101 (2001).

\bibitem{son-etal} W. Son, M.S. Kim, J. Lee and D. Ahn, \jmo{49}, 1739 (2002).

\bibitem{daffer-etal} S. Daffer, K. Wodkiewicz, J. K. McIver,
\pra{67}, 062312 (2003).

\bibitem{diosi} L.  Diosi, in {\em Irreversible Quantum Dynamics},
edited by F. Benatti,  R. Floreanini (Springer, New York, 2003),
pp. 157-163.

\bibitem {dodd-halliwell} P. J. Dodd and J. J. Halliwell, \pra
{69}, 052105 (2004).

\bibitem{yu-eberly09} T. Yu and J. H. Eberly, \sci {323}, 598 (2009).

\bibitem{qian-eberly12} Q.F. Qian and J.H. Eberly, Phys. Lett. A {\bf 376}, 2931 (2012).

\bibitem{QEC} See, for example, P.W. Shor, \pra{52}, R2493-R2496 (1995); A. M.  Steane, \prl{77}, 793 (1996); C. H. Bennett, D.P. DiVincenzo, J.A. Smolin, and W.K. Wootters, \pra{54}, 3824 (1996).

\bibitem{wickert-etal} R. Wickert, N.K. Bernardes and P. v. Loock, \arx{1004.1931} (2010).

\bibitem{shor} P. W. Shor, \pra{52}, R2493-R2496 (1995).

\bibitem{gottesman} D.  Gottesman, \emph{Stabilizer Codes and Quantum Error Correction}, PhD thesis, California Institute of Technology, Pasedena, CA, USA, 1997. 

\bibitem{carmichael} H. Carmichael, \emph{Statistical Methods in Quantum Optics: Master Equations and Fokker-Planck Equations} (Springer, Germany, 2002), p.  6.

\bibitem{wootters} W. K. Wootters, \prl {80}, 2245 (1998).

\bibitem{laflamme} R.  Laflamme, C.  Miquel,  J.  P.  Paz, and W.  H.  Zurek, \prl{77}, 198 (1996).

\bibitem{yu-eberly06}  T.Yu and J.H. Eberly, \oc {264}, 393 (2006).

\bibitem{yu-eberly10} T.Yu and J.H. Eberly, \oc {283}, 676 (2010).

\bibitem{leung} D.  W.  Leung, M.  A.  Nielsen, I.  L.  Chuang, and Y.  Yamamoto, \pra {56}, 2567 (1997).

\bibitem{chaves}  R. Chaves, L. Aolita and A. Ac\'{\i}n, \pra{86}, 020301(R) (2012).  

\end{thebibliography}
\end{document}